# News Harvesting from Google News combining Web Scraping, LLM Metadata Extraction and SCImago Media Rankings enrichment: a case study of IFMIF-DONES

Víctor Herrero-Solana

SCImago-UGR, Unit for Computational Humanities and Social Sciences (U^CHASS), University of Granada, Spain

https://orcid.org/0000-0003-1142-5074

## Abstract

Objetive. This study evaluates Google News as a complementary data source for communication research, documenting its capabilities and limitations through a case study implementing automated web scraping, large language model (LLM)-based metadata extraction, and SCImago Media Rankings enrichment.

Methodology. Using the IFMIF-DONES fusion energy project as a case study, we implemented a five-stage data collection pipeline across 81 region-language combinations, yielding 1,482 validated records after a 56% noise reduction. Results are compared against two licensed press databases: MyNews (2,280 records) and ProQuest Newsstream Collection (148 records). Results. Overlap analysis reveals high complementarity, with 76% of Google News records exclusive to this platform. The dataset captures content types absent from proprietary databases, including specialized outlets, institutional communications, and social media posts. However, significant methodological challenges emerge: temporal instability requiring synchronic collection, a 100-result cap per query demanding multi-stage strategies, and unexpected noise including academic PDFs, false positives, and pornographic content infiltrating results through black hat SEO techniques. LLM-assisted extraction proved effective for structured articles but exhibited systematic hallucination patterns requiring validation protocols.

Conclusions. We conclude that Google News offers valuable complementary coverage for communication research but demands substantial methodological investment, multi-source triangulation, and robust filtering mechanisms to ensure dataset integrity.

## 1. Introduction

Media content analysis has long relied on systematic news collection as its empirical foundation. Traditionally, researchers have depended on subscription-based databases such as LexisNexis, Factiva, ProQuest, and specialized regional services to access news archives for content analysis studies (Weaver & Bimber, 2008; Ridout et al., 2012). These professional databases offer structured search capabilities, consistent metadata, and longitudinal coverage spanning decades. However, they present significant limitations: geographic and linguistic biases favoring English-language and Western media outlets, substantial subscription costs that may be prohibitive for researchers in resource-limited contexts, and restricted coverage of digital-native and regional news sources that increasingly dominate contemporary media landscapes.

The emergence of online news aggregators, particularly Google News, has fundamentally transformed the accessibility of news content for research purposes. These platforms aggregate content from tens of thousands of sources across multiple languages and regions, offering researchers unprecedented breadth of coverage without subscription barriers. Google News alone indexes content from over 50,000 publishers in 35 languages (Newman et al., 2024), although Google does not provide any specific listing and maintains the portal as the "black-box" of world news. This democratization of access has made news aggregators increasingly attractive for communication researchers seeking comprehensive media coverage of specific topics, events, or phenomena.

Nevertheless, the use of news aggregators for systematic research introduces methodological challenges that warrant careful consideration. Unlike curated databases with documented source selection criteria, aggregator algorithms operate as opaque gatekeepers whose selection and ranking mechanisms remain largely undisclosed (Nechushtai & Lewis, 2019). Research has demonstrated that Google News exhibits structural biases in source representation, over-representing certain mainstream outlets while under-representing others, regardless of personalization effects (Haim et al., 2018). Furthermore, technical constraints limit result retrieval to approximately 100 items per search query, necessitating strategic approaches to maximize coverage through temporal fragmentation and query variation.

This study addresses these methodological tensions by evaluating Google News as a data source for systematic news collection through a documented case study. We assess the platform's capabilities and limitations, implementing an automated pipeline combining web scraping with artificial intelligence-based metadata extraction. The approach responds to calls in computational journalism scholarship for transparent, replicable methods that can accommodate the scale and heterogeneity of contemporary digital news ecosystems (Baden et al., 2022; Brown et al., 2025).

## 2. Case Study: IFMIF-DONES

The IFMIF-DONES project (International Fusion Materials Irradiation Facility – Demo Oriented NEutron Source) is a major international scientific infrastructure currently under construction in Escúzar, Granada, Spain. This neutron source, based on a high-current particle accelerator, is designed to qualify materials capable of withstanding the extreme conditions expected in future fusion reactors (Bernardi et al., 2022; Królas et al., 2021). The facility forms a central element of the European Roadmap to Fusion Energy and has been designated as a strategic research infrastructure by ESFRI since 2018 (Ibarra et al., 2019).

With an initial budget of approximately 700 million euros, funding sources include the European Regional Development Fund (ERDF), EUROfusion, and various European and national programs (Esteban et al., 2021). The project's international dimension was further consolidated in November 2025 with the signing of the DONES Multilateral International Agreement (MIDA), which formalized the participation of Croatia, Japan, and Italy alongside Fusion for Energy (F4E), establishing a comprehensive framework for governance and coordinated development of fusion technologies.

Previous research on IFMIF-DONES has examined public perception through direct surveys among residents in the surrounding area, finding that perceived safety correlates positively with expectations of economic impact (Troya et al., 2022). The present study aims to develop and evaluate data sources for building a comprehensive news dataset that will enable systematic media content analysis of the project. Such analysis would provide an alternative measure of public interest and institutional visibility based on media presence. The characteristics of IFMIF-DONES—international scope, technical complexity, regional economic impact, and positioning within energy policy debates—make it an ideal case for evaluating automated news collection methodologies.

## 3. Objectives

The objective of this study is to evaluate Google News as a data source for media content analysis, using the IFMIF-DONES project as a case study. To this end, we address the following research questions:

- RQ1: To what extent can Google News support the construction of a comprehensive and replicable news dataset?
- RQ2: To what extent do news results in Google News overlap with the indexed content of licensed press databases like MyNews and Newsstream Collection?
- RQ3: How feasible is data extraction from diverse news outlets? Can LLMs assist in this process?
- RQ4: How does the source ecosystem captured by Google News differ from MyNews and Newsstream Collection in outlet diversity, specialization, and agency coverage?
- RQ5: Does Google News introduce noise in the dataset construction process?

## 4. Materials and Methods

This study adopts a single case study design with embedded multiple units of analysis, following the methodological framework established by Yin (2018). According to this author, case study research constitutes an empirical investigation that examines a contemporary phenomenon within its real-world context, being particularly appropriate when the boundaries between the phenomenon and its context are not clearly evident. The IFMIF-DONES project presents these characteristics: media coverage of a scientific infrastructure is inextricably linked to broader contexts of energy policy, regional economic development, and public perception of nuclear technologies. While the case remains singular (IFMIF-DONES media coverage), the research incorporates three distinct data sources as embedded units enabling systematic comparison of collection methodologies, source coverage, and outlet characteristics. This embedded design follows Yin's (2018) recommendation for enhancing analytical depth within single cases. Additionally, as Eisenhardt (1989) notes, case-based research is particularly suitable for novel thematic areas, as it allows generating empirically grounded insights from systematic analysis.

The combination of automated news aggregator scraping with LLM-based metadata extraction represents an emerging area where established protocols are still emerging. While findings from a single case cannot be generalized statistically, they enable analytical generalization (Yin, 2018): the multi-stage data collection pipeline developed here can be transferred and adapted to other scientific infrastructure projects or topics requiring comprehensive media coverage analysis across multiple sources, languages, and countries.

### 4.1 Data Sources

We had access to the Spanish MyNews database, which has been used in content analysis studies in several topics. However, the media coverage in this database is heavily biased toward Spanish outlets; although it includes 139 foreign media sources, this limitation necessitated complementary international sources.

First, we incorporated a database collection from the ProQuest portal known as Newsstream Collection (US+Canadian+International), which includes news from approximately 2,000 media outlets worldwide. This collection is considered one of the most comprehensive globally by number of titles, comparable to the well-known Factiva (Gilbert, 2024). It has been widely used in the literature for content analyses across diverse fields, particularly health-related topics such as nutrition.

The second complementary source selected was Google News, one of the most widely used news aggregation platforms globally and extensively examined in scholarly literature (Lopezosa, 2024). Foundational methodological work by Weaver and Bimber (2008) demonstrated that Google News retrieves substantially more unique news stories than traditional databases like LexisNexis, particularly for topics with broad international coverage. Furthermore, Google News appears to be particularly well-received among younger audiences as a news access channel (Lee, 2015) and, unlike social media platforms, does not appear to produce the filter bubble effect commonly associated with algorithmic content curation (Cardenal, 2019; Bonart, 2020). A notable advantage over traditional news databases is the preservation of news photographs, which are typically excluded from text-based databases (Weaver & Bimber, 2008).

The platform's accessibility, breadth of source coverage, and inclusion of digital-native outlets, regional media, and non-English publications (Watanabe, 2013) that remain underrepresented in traditional subscription databases make it particularly valuable for international scientific infrastructure projects like IFMIF-DONES, where media coverage spans multiple countries, languages, and media ecosystems. While acknowledging the methodological challenges inherent to news aggregators—including algorithmic opacity and source concentration biases (Haim et al., 2018; Nechushtai & Lewis, 2019)—we argue that the unparalleled breadth and accessibility of Google News makes it an indispensable complement to traditional databases for achieving comprehensive media coverage analysis.

### 4.2 Data Collection

Data collection from the two subscription databases was relatively straightforward. Newsstream Collection, operating within the ProQuest portal, allows searching by headline and article body with export to Excel files. MyNews required date-segmented searches to manage data export limitations (Repiso, 2016).

The construction of the Google News dataset, however, proved considerably more complex. The first issue encountered is that regardless of the search query, the portal typically returns no more than 100 results. Indeed, examining the literature that employs this portal reveals studies using various strategies to minimize the impact of this limitation: downloading only one article per day (Ungar, 2008), using a Chrome plugin (Puschmann, 2019), or analyzing a broad topic such as China but retaining only 40 articles (Hung, 2023).

The approach includes three steps: (1) news article discovery through the aggregator platform, (2) full-text retrieval from original sources, and (3) metadata enrichment using artificial intelligence and an open media ranking. This computational approach responds to the growing need for scalable methods in journalism and media studies, particularly given the vast volume of digital news production that exceeds traditional manual coding capabilities (Boumans & Trilling, 2016; Günther & Quandt, 2016). This approach addresses the need for documented, transparent methods in computational journalism research (Baden et al., 2022).

Contemporary news aggregators such as Google News employ dynamic JavaScript rendering, which necessitates browser automation rather than conventional HTTP request methods. Following established practices in computational communication research (Freelon, 2018; Landers et al., 2016), we utilized Selenium WebDriver to simulate user browsing behavior and capture dynamically loaded content. The collection instrument incorporated several design features to ensure systematic and reliable data gathering: randomized request intervals (1.5–4.0 seconds) to prevent server overload, User-Agent rotation to simulate diverse browser environments, and configurable parameters for temporal scope and geographic filtering. These technical considerations align with methodological standards for web-based data collection in communication research (Krotov & Silva, 2018; Brown et al., 2025). Additionally, to avoid explicit user bias, we operated without being logged in to prevent working with Spain's default portal version.

Google News currently supports 38 different languages, has over 60 region-specific versions, and identifies up to 240 different country codes. In total, valid region+language combinations number 81. Using Claude/Opus 4.5, we generated 81 different Python scripts to query the portal. Knowing in advance that the Spanish+Spain version (es:ES) would yield the most results, we launched scripts with timeframes covering single months from January 2016 to December 2025 (after:XXXX before:XXXX), totaling 120 scripts. For all other country versions, simple searches were conducted across all of them (site:ISO), totaling 89 scripts. Subsequently, all articles retrieved up to that point were collected, and outlet domains were used to launch new searches (site:domain) on the two portals with the highest previous results: es:ES and en:US. In total, this comprised 264 domains × 2 portals = 528 scripts. All resulting files were merged and duplicates removed. Duplicate detection combined URL matching with headline similarity analysis to ensure unique records. For each discovered article, the system resolved redirect

URLs to access original publisher content, thereby preserving source attribution, a critical consideration for media research (Karlsson & Sjøvaag, 2016). Article text was extracted using HTML parsing techniques that prioritized semantic content elements while filtering navigational and advertising components.

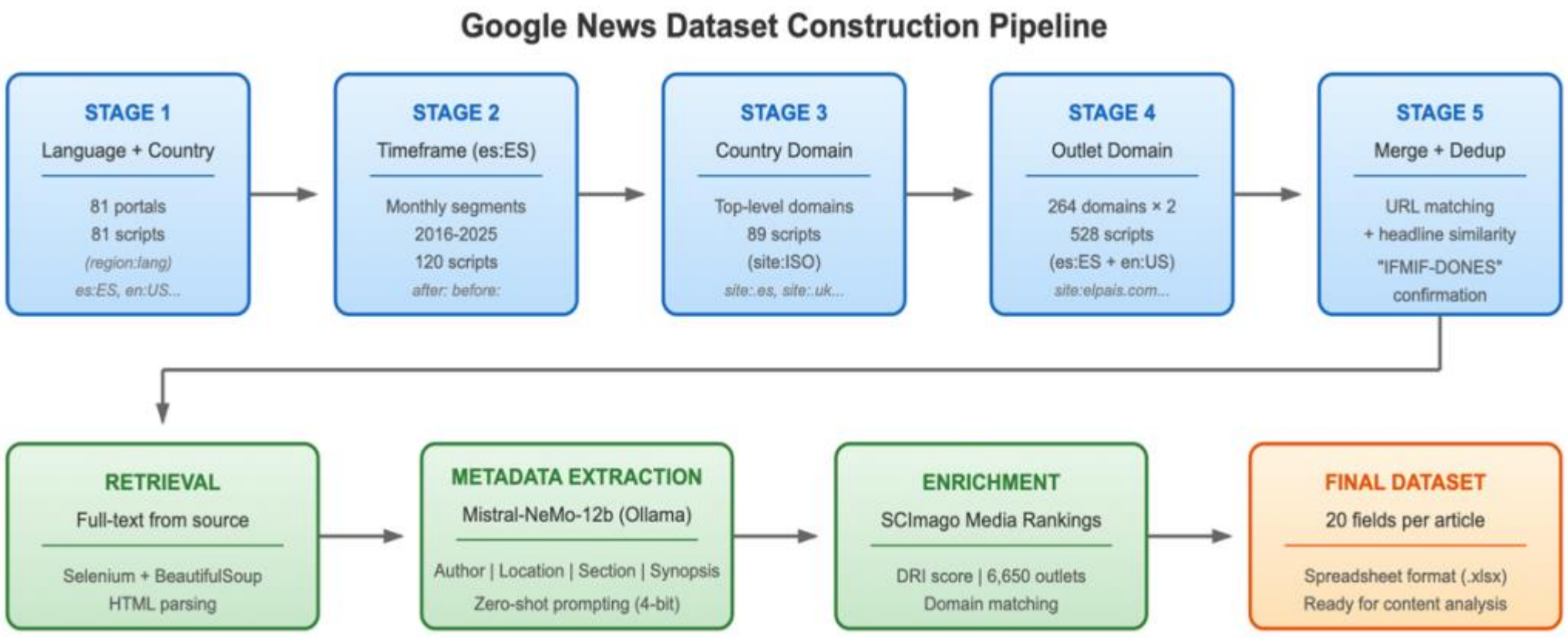

Figure 1 – Google News dataset construction pipeline

The pipeline incorporates large language models (LLMs) for automated metadata extraction, following recent research demonstrating that LLMs achieve accuracy comparable to trained human coders in structured information extraction tasks (Turner et al., 2025; Gilardi et al., 2023). We employed Mistral-NeMo-12b, an open-weights language model, running locally to ensure data privacy and eliminate API costs. Previous experience with 25 open models for gender inference indicated that this model was among the most efficient relative to its size (Herrero-Solana et al., 2026). We selected the Ollama portal version with 4-bit quantization, which is sufficient for data identification tasks. The model extracted four metadata categories using zero-shot prompting: authorship, geographic focus, thematic section, and synopsis. This AI-assisted approach substantially reduces the manual coding burden while maintaining methodological rigor, addressing a persistent challenge in large-scale media content analysis (Baden et al., 2022). The resulting dataset was structured in spreadsheet format (.xlsx) containing 20 fields per article: unique identifier, outlet name, source domain, publication timestamp, collection timestamp, headline, author, geographic reference, thematic category, source URL, full text, character count, word count, featured image URL, extraction method indicator, detected language, AI-generated summary, and data quality flags. For articles downloaded without content, a new script was created to attempt completing the content field through a specific new query. Here, the script had to handle problems with paywalls, PDF formats, video, audio, and similar issues.

While Mistral-NeMo-12b demonstrated high accuracy in metadata extraction for well-structured articles, we observed a specific hallucination pattern during full-text retrieval. When technical barriers prevented content access, due to paywalls, JavaScript rendering failures, server timeouts, or anti-bot protection, the model generated fabricated text rather than returning an empty field or error message. Notably, this hallucinated content was identical across all affected records, producing a consistent placeholder text regardless of the source article. This uniformity enabled straightforward detection through simple string matching, allowing systematic identification and removal of all affected entries. Records flagged with hallucinated content were either reprocessed manually. This experience underscores that while LLM-assisted extraction substantially reduces manual coding burden, validation mechanisms remain essential, though certain hallucination patterns may exhibit regularities that facilitate automated quality control.

To assess extraction accuracy, we conducted a preliminary validation on a stratified random sample of 50 records, comparing LLM-extracted metadata against actual page content (Table 1). The sample was proportionally stratified by source typology. Of the 50 URLs, 46 (92%) remained fully accessible and 4 (8%) were paywalled. Title extraction achieved 100% exact match across all 50 records. Author identification proved more challenging: of the 15 records where authorship could be evaluated, 6 were correctly extracted, 2 were incorrect, and 7 were empty despite bylines being present on the page. The remaining 35 records had no identifiable author on the source page. Content validity was confirmed in all 47 evaluable records (44 fully valid, 3 partially valid), with zero hallucinations detected in the final dataset. These results suggest reliable extraction for structured fields (titles, body text) but limited capacity for less standardized metadata such as authorship, consistent with the heterogeneous structure of sources in the dataset.

| Field | Evaluable | Exact | Wrong | Empty | N/A | Accuracy |
|---|---|---|---|---|---|---|
| Title | 50 | 50 | 0 | 0 | 0 | 100% |
| Author | 15 | 6 | 2 | 7 | 35 | 40.0% |
| Content* | 47 | 44 | 0 | - | 3† | 93.6% |

* Content categories: Valid/Partial/Hallucinated (44/3/0). † Inaccessible.

Table 1 - Preliminary validation of LLM-extracted metadata (n = 50)

The final step involved enriching each outlet with complementary information. For this purpose, we utilized SCImago Media Rankings, an open-access portal providing information on 6,650 outlets worldwide, including name, domain, country, region, language, typology, and a reputation indicator (Trillo-Domínguez et al., 2023). The Digital Reputation Indicator (DRI) is a composite webometric index designed to assess the digital reputation of news media through objective visibility and linking metrics. It combines four standardized indicators obtained from globally recognized SEO platforms: Authority Score (SEMrush), Domain Rating (Ahrefs), Citation Flow and Trust Flow (Majestic). These indicators capture both popularity (volume of incoming links) and prestige (quality and authority of linking domains). All metrics are normalized and equally weighted to avoid source-specific bias. The final DRI score is calculated as the aggregated mean of the four indicators, enabling systematic, comparable, and scalable rankings of digital media across countries and languages based on their web-based reputation rather than audience size or social media engagement (Trillo-Domínguez et al., 2025). Some authors have begun to use it to dispense with Google News portals and thus scrape directly a geographically defined domain of websites, such as Spanish-speaking countries (Barredo-Ibáñez & Barrera-Jerez, 2026). This information serves not only to identify when an article comes from a recognized outlet but also to determine its reputation level.

Data collection and processing were conducted on consumer-grade hardware (Apple Mac Mini M4, 24GB unified memory) running Python 3.11, Selenium 4.x, BeautifulSoup 4.x, and the Ollama framework hosting Mistral-NeMo-12b (4-bit quantization, ~5GB memory). This configuration demonstrates the accessibility of computational methods for communication researchers without requiring specialized infrastructure. The complete source code and datasets are available at https://doi.org/10.5281/zenodo.18616551 to support methodological transparency and replication (Dienlin et al., 2021).

## 5. Results

The final datasets comprised: MyNews with 2,280 records, Google News with 1,482 records, and International Newsstream Collection with 148 records. Before examining each dataset in detail, it is important to describe how Google News records were progressively accumulated (and eliminated), as illustrated in Figure 2. In Stage 1, the initial search yielded 425 records. At this point, we observed that only two portals contributed significant results: Spain (es:ES) and the United States (en:US). Other Spanish-language portals largely replicated results from the Spanish version to varying degrees. The 9 Spanish-language portals produced 854 records, which after deduplication yielded only 157 unique entries. For English, 21 portals generated 378 records, but only 22 were unique. Results from other languages proved negligible.

In Stage 2, applying the timeframe segmentation technique produced additional volume (594 records), but most were duplicates, raising the cumulative total to only 524 unique records. Utilizing the domain parameter significantly increased yield. Stage 3 retrieved 231 Spanish-language records (143 unique) and 330 English-language records (242 unique). At this point, we began detecting problems with results quality, particularly the proliferation of irrelevant content, as discussed in the following section. The cumulative total reached 1,160 records.

Stage 4 contributed substantially more. English searches yielded 960 records (678 unique), while Spanish searches produced 2,824 records (2,452 unique). We initially believed we had amassed a news volume considerably larger than MyNews, having reached 4,277 records (3,340 unique). However, substantial noise remained. Much of the material consisted of non-news content or was simply off-topic. To filter these, we systematically searched for the string "IFMIF-DONES" throughout the dataset, identifying numerous entries that lacked this term entirely. Others pointed to non-HTML resources: PDFs, images, audio, video, and even executables. The final filtered dataset contained only 1,482 records, representing a 56% reduction from the pre-filtering total.

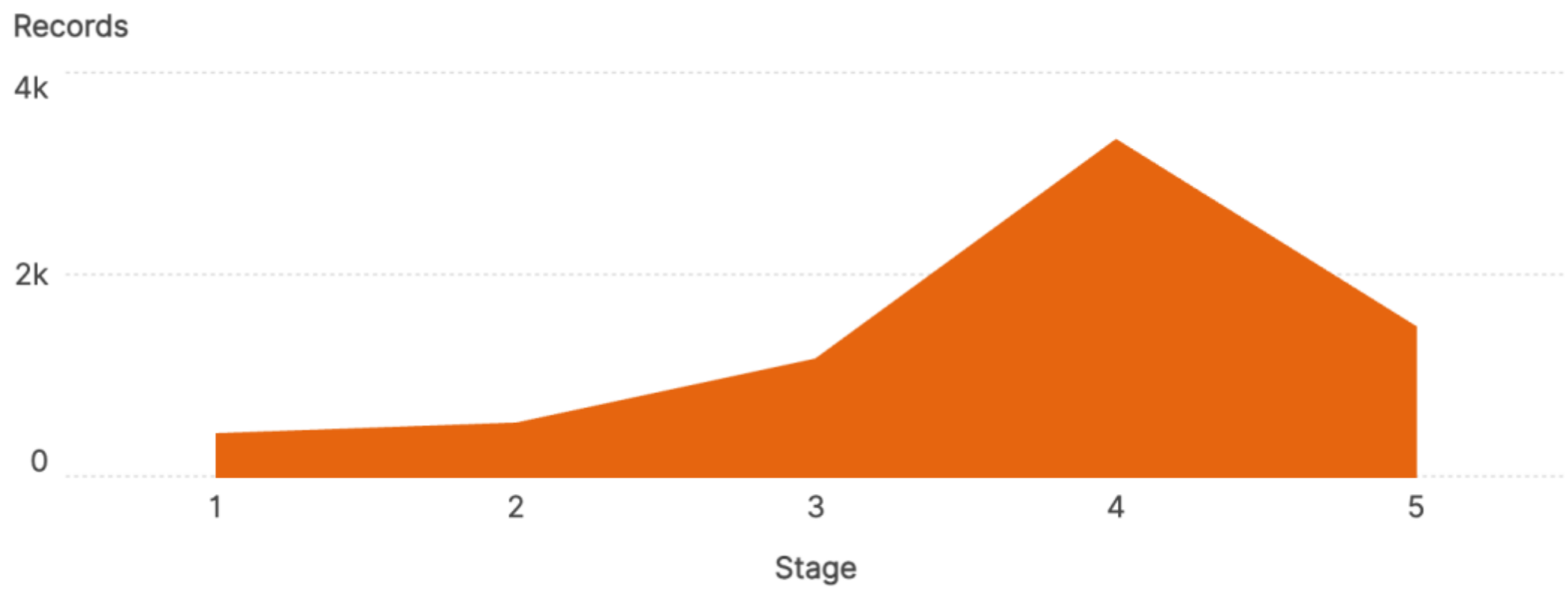

Figure 2 – Evolution of the Google News dataset construction

The overlap among the three datasets is proportionally low, with each source contributing a substantial number of exclusive records. Exclusive content percentages are 70% for Newsstream Collection, 76% for Google News, and 84% for MyNews. This demonstrates that no source is subsumed within another; each contributes original information and they are therefore highly complementary. This finding validates the multi-source approach adopted in this methodology.

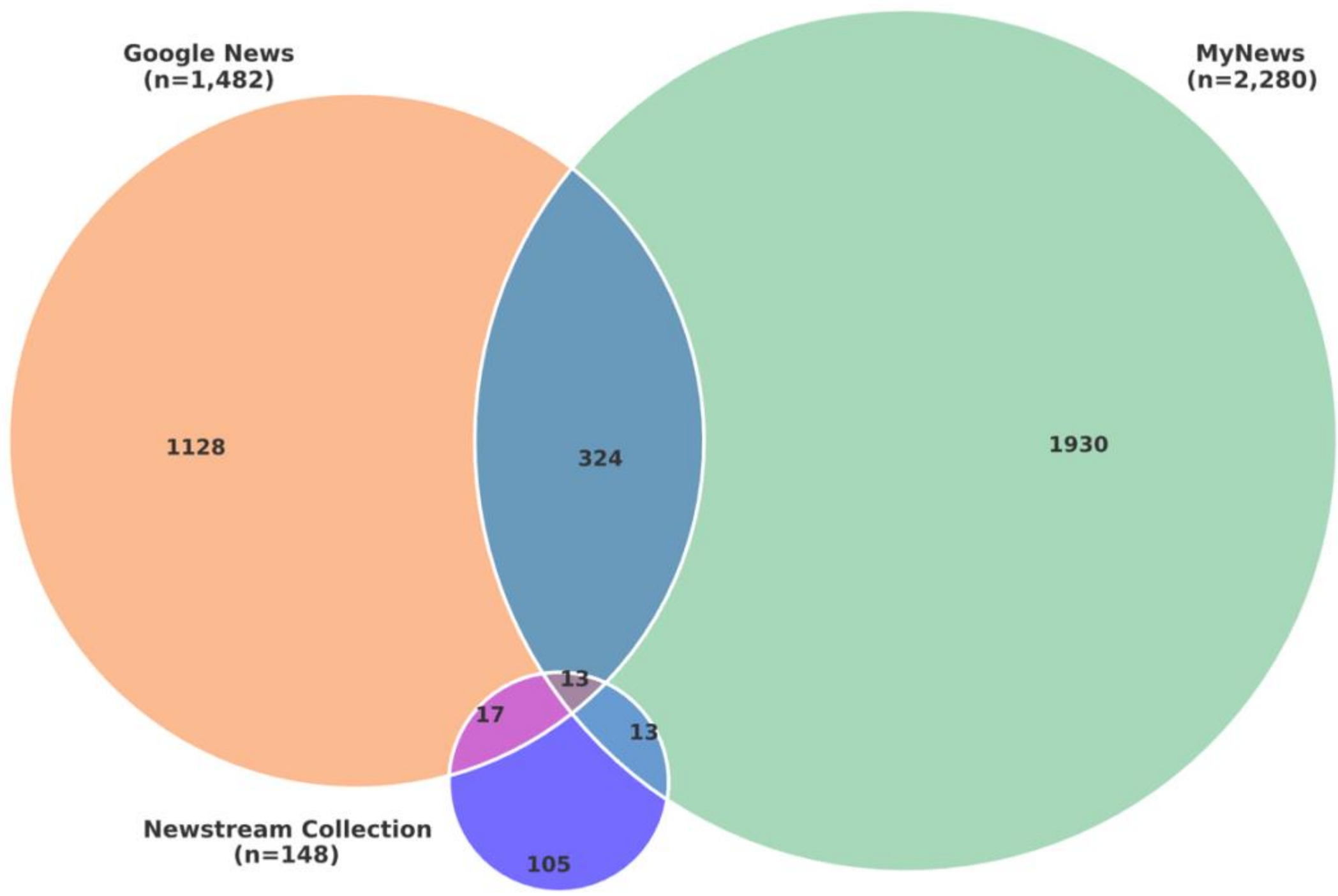

Figure 3 – Sources overlap

Regarding geographic origin of coverage, our a priori assumption that MyNews would be predominantly Spanish was confirmed. Its international component is comparatively small, comprising outlets from only two countries: Argentina and Italy. Moreover, the Argentine content is predominantly from Infobae, a Buenos Aires-based outlet that has expanded across several Spanish-speaking countries including Spain itself, and could therefore be considered quasi-national press. The converse also applies: we classified all El País content as Spanish, although some articles may have been published in its Latin American editions.

Newsstream Collection presents international news from seven different countries, with international volume equivalent to its Spanish media coverage. The most pronounced difference appears in Google News, which presents the largest volume of international information from over thirty different countries. While many countries are represented, most contributions are modest, none exceeding ten articles. The top five contributing countries are: Argentina (9), Croatia (8), United Kingdom (8), Italy (7), and Mexico (6).

However, another variable warrants consideration: outlet inclusion in the SMR (Figure 4). Here we observe clearly that this international content volume predominantly originates from sites not included in SMR. Both MyNews and Newsstream Collection present substantially higher percentages of SMR-listed outlets compared to Google News. While this is not necessarily negative, it merits deeper examination. As discussed below, the nature of this non-SMR content is highly diverse.

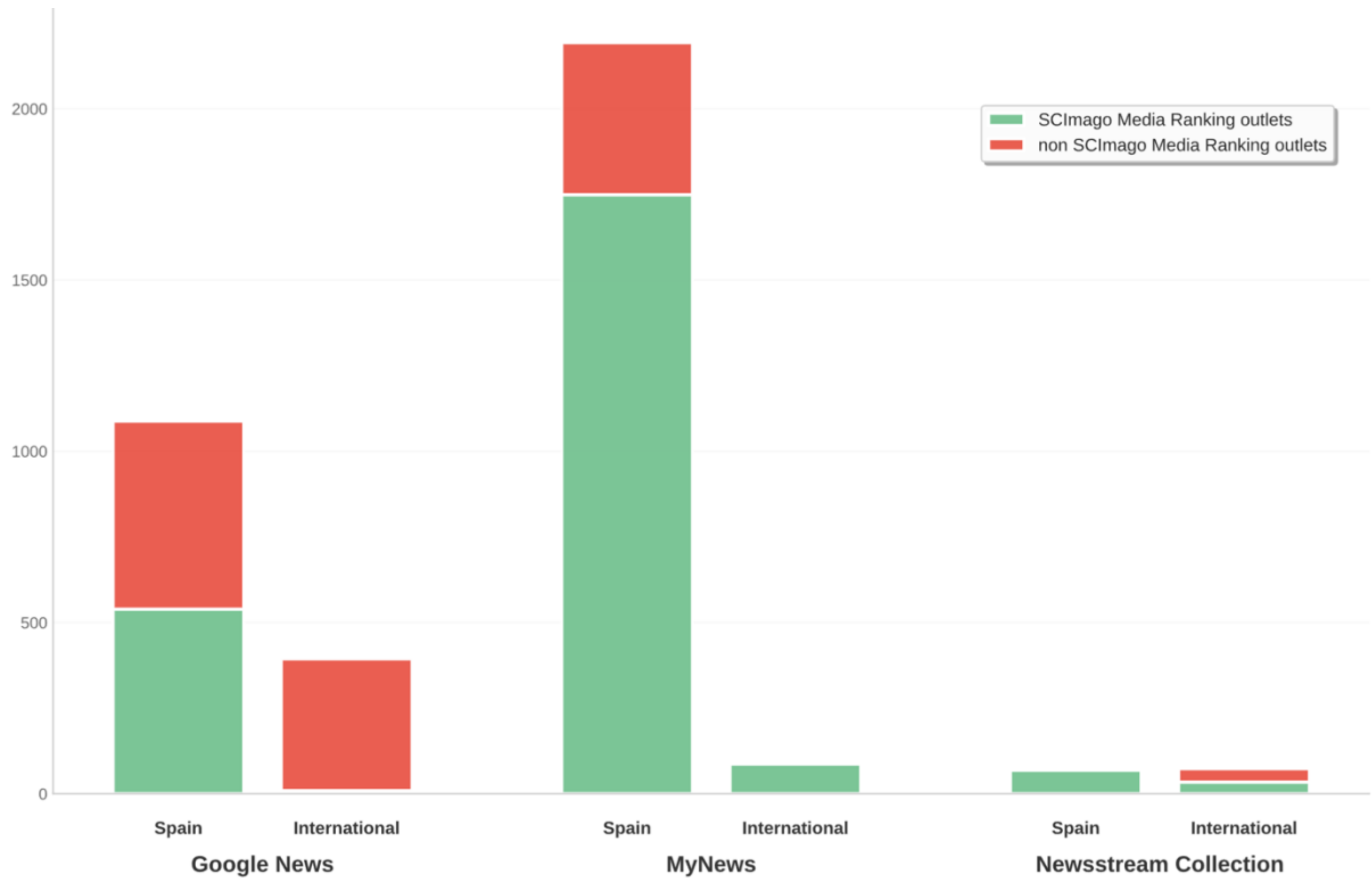

Figure 4 – Sources by country and SMR status

Figure 5 presents a box-and-whisker plot of the Overall field values from SMR-listed outlets for articles in each dataset. The dispersion patterns for Google News and MyNews are quite similar, both for Spanish and international news—although the latter visualization is less informative given the small number of cases. A different pattern emerges for Newsstream Collection, whose mean lies above the other two sources, though whiskers are not displayed due to the limited number of represented elements. This suggests that Newsstream Collection, while smaller in volume, may index outlets with higher average digital reputation.

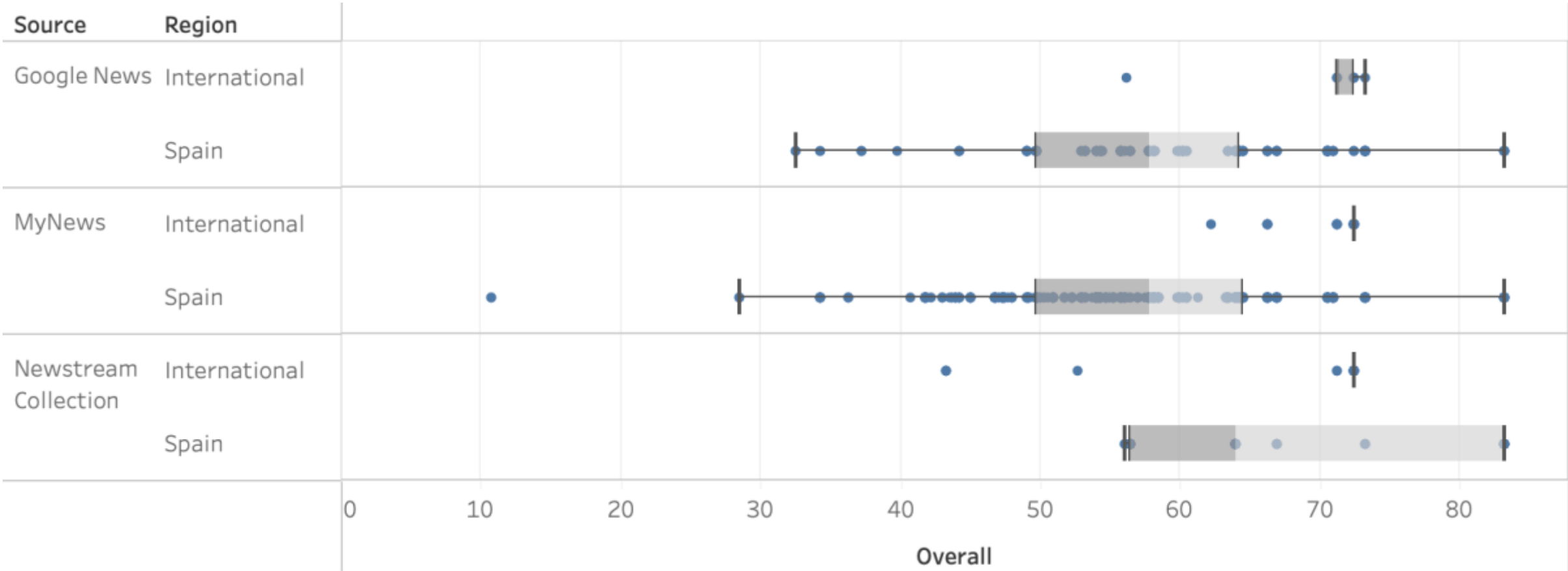

Figure 5 – Overall values from SCImago Media Rankings (SMR)

By their nature as journalistic information sources, non-SMR elements in MyNews and Newsstream Collection are likewise media outlets that simply have not entered the ranking—either because the outlet is very small (typically a digital-native newspaper) or because it represents radio and television media, which will soon be incorporated into the portal. The nature of Google News non-SMR results differs substantially; while it captures small outlets, it also clearly extends into other domains. Figure 6 presents a treemap visualization of the dataset portion that did not match with SMR.

The largest segment, shown in blue, comprises other outlets not in SMR (360 records). Several of these media also appear in the non-SMR segment of MyNews, consisting primarily of small digital-native media from Granada, radio, and television. The most prominent is elindependientedegranada.es, a hyperlocal digital newspaper. However, we also identified more specialized outlets covering technology (xataka.com), science (novaciencia.es), innovation (innovaspain.com), energy (elperiodicodelaenergia.com), and nuclear fusion (world-nuclear-news.org).

The second segment comprises social network content, shown in yellow. These include major platforms: LinkedIn, X (formerly Twitter), Facebook, YouTube, and Instagram. While some posts originate from media outlets, others come from companies, universities, and individuals, leading us to present them separately. This segment comprises approximately 300 records. The third segment, shown in red, represents government sources (114 records). This includes administrations at multiple levels: Spain (national), Andalusia (regional), and Granada (local). Content typically originates from public administration press offices. Political parties are also included in this category. Some international content appears, centered on countries with concrete interests in the project: Italy, Croatia and Japan.

Approximately equivalent in size (106 records) is the segment designated Science & Technology (S&T) Institutions. This includes universities, research centers, and scientific societies. The University of Granada features most prominently, as the local S&T institution most engaged with the project. Of the three remaining segments, perhaps most significant is transnational projects, which includes the DONES initiative itself. Despite this centrality, the number of records found is relatively small, suggesting that project-generated content may be underrepresented in news aggregator indexing.

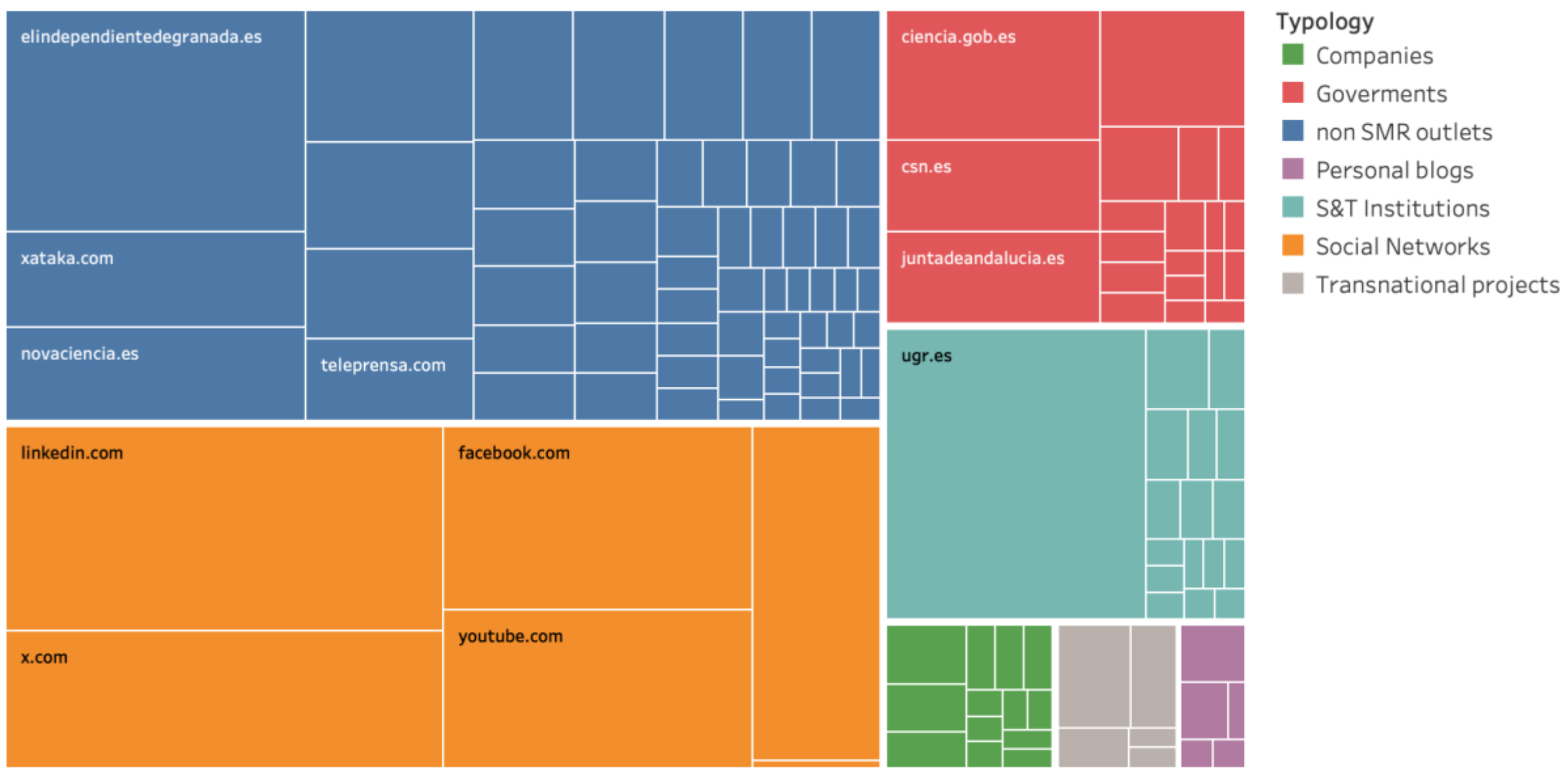

Figure 6 – non SMR sites typology of Google News dataset

## 6. Discussion

- RQ1: To what extent can Google News support the construction of a comprehensive and replicable news dataset?

Addressing this question requires unpacking the concept of "robustness" into three constituent dimensions: volume adequacy (whether sufficient records can be retrieved), content validity (whether retrieved content represents genuine news coverage), and methodological transparency (whether the collection process can be documented and reproduced).

Regarding volume, our five-stage pipeline ultimately yielded 1,482 validated news records about IFMIF-DONES, representing approximately 65% of the volume obtained from MyNews (2,280 records) and substantially exceeding Newsstream Collection (148 records). This volume demonstrates that Google News can generate datasets of sufficient scale for meaningful content analysis. However, achieving this required substantial methodological investment: 737 distinct scripts executing searches across 81 region-language combinations, five sequential collection stages, and extensive post-processing that reduced an initial pool of 4,277 candidate records by 56% after validation filtering.

Content validity presented greater challenges. The necessity of synchronic collection has been documented in prior research. Prior research confirms this temporal instability: Herrero-Solana et al. (2014) required polling intervals as short as 15 minutes and over 1.5 million queries across a year-long study to mitigate data loss. Our findings align with these observations: URL decay rates proved substantial, with many links becoming inaccessible within weeks of initial indexing. This temporal instability renders Google News largely unsuitable for retrospective data collection, where researchers seek to reconstruct historical coverage of past events.

Methodological transparency—a cornerstone of replicable research—faces fundamental obstacles with Google News. Unlike licensed databases that maintain stable archives and documented indexing criteria, Google News operates as a dynamic aggregator whose selection algorithms remain opaque and whose results vary based on query timing, user location, and personalization factors. Karstens et al. (2023) showed that even Factiva and Nexis Uni return different results for identical searches at different times, though variability in Google News substantially exceeds that of licensed databases. Buntain et al. (2023) identified additional limitations including result caps and algorithmic personalization introducing systematic bias.

These findings suggest that Google News offers moderate support for constructing comprehensive datasets but proves inadequate as a sole data source for rigorous media research. Replicability requires exhaustive documentation of collection parameters, timestamps, and filtering criteria—a methodological burden that researchers must anticipate and resource accordingly. The platform is better suited for prospective monitoring designs, where real-time collection can mitigate temporal instability, than for retrospective studies requiring historical coverage reconstruction.

- RQ2: How does Google News coverage overlap with licensed press databases, and what does this imply for multi-source research designs?

The overlap between Google News and the licensed news databases is approximately 30%, demonstrating that Google News serves as a valuable complement rather than a substitute for traditional databases. Each source contributes substantial unique content: 70% of Newsstream Collection records, 76% of Google News records, and 84% of MyNews records are exclusive to their respective sources.

This low overlap aligns with findings from previous comparative studies. Weaver and Bimber (2008) reported that Google News retrieved substantially more unique news stories than LexisNexis, particularly for topics with broad international coverage—a pattern consistent with our results, where Google News contributes the largest volume of international sources. Similarly, Gilbert et al. (2024), comparing six news aggregator databases, found considerable differences in geographic reach and source coverage across platforms, reinforcing that no single database provides comprehensive representation.

The exclusivity asymmetry is notable: MyNews (84%) captures Spanish regional press rarely indexed elsewhere, Newsstream Collection (70%) focuses on mainstream international outlets, and Google News (76%) contributes digital-native outlets, social media, and institutional sources outside traditional media ecosystems.

These findings validate multi-source triangulation and suggest that comprehensive media coverage analysis—particularly for topics spanning multiple countries and media types—requires systematic integration of complementary databases rather than reliance on any single source.

- RQ3: How feasible is data extraction from diverse news outlets? Can LLMs assist in this process?

Data extraction from heterogeneous news sources presents substantial technical challenges that vary considerably across outlet types. The integration of Mistral-NeMo-12b for metadata extraction proved largely successful for structured information identification. However, we observed a specific hallucination pattern during full-text retrieval: when technical barriers prevented access to article content, the model generated fabricated text to fill the empty content field. Crucially, this hallucinated content was consistently identical across all affected records, producing a recognizable placeholder text rather than varied fabrications. This uniformity made detection straightforward: a simple string-matching filter identified and flagged all instances, which were subsequently reprocessed manually. This suggests that certain hallucination patterns may exhibit systematic regularities facilitating automated detection, though further investigation across models is needed.

Selenium WebDriver, while effective for JavaScript-rendered content, exhibited limitations that constrained collection efficiency. Rate limiting and anti-bot measures necessitated conservative request intervals, substantially extending collection timelines. These challenges align with broader assessments of traditional browser automation tools, which note that Selenium's architecture makes it unsuitable for high-volume data collection due to slower command execution and maintenance burden.

We have begun preliminary work with Cowork, Anthropic's agentic desktop automation tool, which demonstrates improved performance over traditional WebDriver approaches. Cowork's ability to interpret page content contextually and adapt to layout variations reduces the brittleness inherent in selector-based automation. Initial results suggest that agentic browser tools represent a promising direction for large-scale news collection, potentially enabling more comprehensive and maintainable data harvesting pipelines for communication research. A systematic evaluation of Cowork's capabilities for automated news harvesting constitutes our next line of investigation.

- RQ4: How does the source ecosystem captured by Google News differ from licensed databases in outlet diversity, specialization, and agency coverage?

Our results reveal a heterogeneous source composition in the Google News dataset that differs markedly from licensed databases. Approximately 37% of records originate from outlets with measurable prestige through the SMR, while 24% come from non-SMR media outlets. This latter category comprises both traditional outlets of limited reach and highly specialized sources—the latter representing Google News's most distinctive contribution, as these specialized outlets rarely appear in proprietary news databases.

Beyond result quantity limitations, research has identified systematic biases in Google News source representation. Haim, Graefe, and Brosius (2018) found that Google News over-represents certain news outlets while under-representing other highly frequented sources, regardless of personalization effects. Similarly, Nechushtai and Lewis (2019) reported that approximately 49% of Google News recommendations came from just five national news organizations, indicating a high degree of source concentration that may not reflect the full diversity of media coverage on any given topic. These findings suggest that researchers cannot assume news aggregator results constitute a representative sample of all available coverage. Our data partially confirm this concentration pattern for mainstream outlets, but simultaneously reveal that Google News captures substantial content from sources entirely absent from traditional databases.

Social media content represents approximately 20% of our Google News dataset, comprising major platforms including LinkedIn, X, Facebook, YouTube, and Instagram. The presence of social media content in Google News results highlights both an opportunity and a methodological consideration: while such content may offer valuable insights into public engagement and institutional communication strategies, it requires distinct analytical treatment from traditional news coverage. Depending on research objectives, this social media component may represent either valuable supplementary data or noise requiring filtration.

Perhaps Google News's most distinctive contribution lies in the heterogeneous category comprising companies, governments, universities, research centers, and project-related sources—representing 18% of our dataset. This segment captures institutional communications from press offices and communication teams that frequently do not reach traditional media outlets. For scientific infrastructure projects like IFMIF-DONES, this institutional content provides direct access to primary source communications that would otherwise remain invisible in conventional media analysis. The diversity within this category precludes straightforward classification but merits positive assessment as a unique data source unavailable through proprietary databases.

Regarding news agency coverage, Weaver and Bimber (2008) demonstrated that LexisNexis is "blind to wire stories," missing half or more of news stories appearing in major newspapers due to wire service exclusions from its indexing. This limitation suggested that Google News might offer superior agency coverage. However, in our specific case, we found no significant differences in agency representation because the MyNews database—unlike LexisNexis—comprehensively indexes agency content, with EFE and Europa Press proving particularly prolific sources for Spanish-language coverage of IFMIF-DONES. This finding underscores that wire service coverage varies substantially across databases, and researchers should verify agency indexing policies when selecting data sources for media analysis.

- RQ5: To what extent does Google News introduce noise in systematic dataset construction, and what forms does this noise take?

This aspect of Google News behavior proved most surprising in our investigation. The multi-stage collection pipeline, while effective at expanding coverage, simultaneously surfaced substantial volumes of content that would typically remain hidden in standard searches—not all of which constituted legitimate news content.

First, we encountered a considerable number of URLs pointing to PDF files (n=221). These records originated predominantly from non-outlet sources: universities (e.g., helmholtz.de, ugr.es, tue.nl, tum.de), institutional repositories (e.g., pure.mpg.de, iris.enea.it, repo.pw.edu.pl), conference proceedings (e.g., jacow.org), international organizations (e.g., iaea.org), multinational laboratories (e.g., cern.ch), and government portals (e.g., hacienda.gob.es, juntadeandalucia.es, mzom.gov.hr, lamoncloa.gob.es). The PDF content comprised scientific papers, technical reports, curricula vitae, official regulations, and procurement documents—material more appropriate for Google Scholar than Google News. This finding suggests that intensive, multi-stage querying on scientific or technical topics forces the emergence of content that remains hidden in conventional searches, effectively blurring the boundary between news aggregation and academic indexing.

Second, we observed the false positive problem previously reported by Hung and Chen (2023), who found that approximately 27% of their Google News results for "China" retrieved the keyword from advertising content rather than article headlines or body text. In our case, numerous URLs pointed not to individual articles but to homepages, section indices, related news listings, or sidebar content. This pattern appeared particularly prevalent among non-SMR outlets such as Gizmodo, Granada Digital, Linares 28, and Teleprensa.com—inflating the apparent result count well beyond the actual number of discrete news articles. Fortunately, this noise was effectively filtered through systematic string-matching for "IFMIF-DONES" within the article body text.

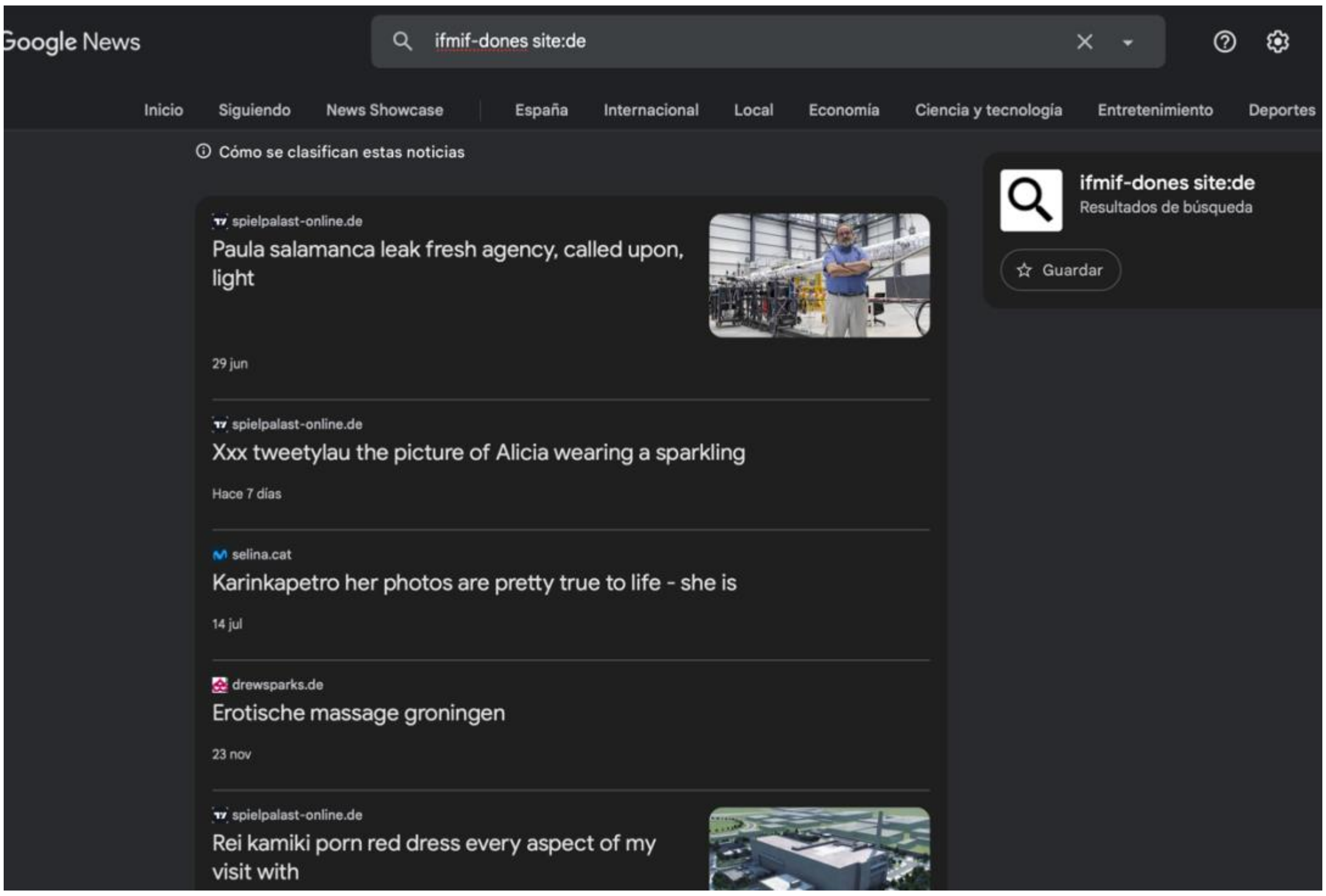

Figure 7 – Porn results in “IFMIF-DONES” in Google News Germany

Most unexpectedly, our dataset included content from pornographic websites (n=20), appearing specifically in results from certain national portal versions (Belgium, France, and Germany). These sites employed black hat SEO techniques—specifically keyword stuffing with trending or technical terms combined with cloaking strategies that present different content to search engine crawlers than to human visitors (Gyöngyi & Garcia-Molina, 2005). While Bevendorff et al. (2024) documented the pervasive problem of SEO spam in Google Web Search, demonstrating that search engines are engaged in an ongoing "cat-and-mouse game" with manipulative content, their analysis focused primarily on commercial affiliate spam rather than adult content infiltration. To our knowledge, the presence of pornographic content in Google News results has not been previously documented in the academic literature, suggesting that the platform's content filtering mechanisms may be more vulnerable than commonly assumed when intensive, multi-stage collection strategies push beyond typical result boundaries. Given the unusual nature of this finding, Figure 7 provides a representative screenshot. Our extraction script automatically flagged these entries as [NO CONTENT] when full-text retrieval failed, enabling straightforward removal from the final dataset.

These findings collectively demonstrate that while Google News can substantially expand news coverage for specialized topics, researchers must implement robust validation and filtering mechanisms. The noise-to-signal ratio increases substantially when collection strategies extend beyond simple keyword searches, requiring systematic quality control procedures that should be anticipated and documented in methodological protocols.

## 7. Conclusions

This study demonstrates that Google News offers valuable complementary coverage for communication research, though its effective use requires substantial methodological investment and careful attention to data quality. Our five-stage pipeline successfully retrieved 1,482 valid news records about IFMIF-DONES, representing 65% of the volume obtained from the leading licensed database (MyNews) while capturing content largely absent from proprietary sources—including specialized outlets, institutional communications, and social media posts that collectively constitute 62% of the Google News dataset.

The complementary value of Google News is clear: approximately 70% overlap absence between databases confirms that no single source provides comprehensive coverage, and each captures distinct segments of the news ecosystem. Google News excels at surfacing diverse outlet types and non-traditional sources that licensed databases systematically exclude, offering researchers access to content that would otherwise remain invisible to academic inquiry.

However, significant limitations constrain its applicability. Retrospective data collection proves largely impractical due to result volatility, URL decay, and the platform's inherent lack of archival stability. Researchers requiring historical news coverage should rely on licensed databases, reserving Google News for prospective, synchronic monitoring where real-time collection can mitigate temporal instability. The platform is better suited as a complementary source within multi-database designs than as a standalone data collection instrument.

Technical implementation presents considerable challenges. Web scraping pipelines using conventional tools such as Selenium WebDriver remain brittle, requiring continuous maintenance as page structures evolve and anti-bot measures intensify. However, LLM-assisted metadata extraction offers promising efficiency gains, enabling automated classification of authorship, geographic focus, and thematic content—provided appropriate validation protocols address hallucination risks through systematic verification procedures. Emerging agentic desktop automation tools may improve on traditional WebDriver approaches, though systematic evaluation for news harvesting remains for future research.

Finally, noise reduction demands particular attention. Intensive collection strategies push beyond Google News's typical result boundaries, surfacing academic PDFs misindexed from Google Scholar, URLs pointing to section pages rather than individual articles, and—most concerning—content from sites employing black hat SEO techniques, including pornographic websites exploiting keyword stuffing and cloaking strategies. Robust filtering mechanisms, including body-text keyword verification and content-type validation, are essential components of any methodological protocol.

**Funding**

This work was carried out by two research projects:

1. “Media Narratives on Nuclear Fusion: Impact and Communication Strategies for the IFMIF-DONES Project” (C-SEJ-353-UGR23), funded under the Applied Research Projects call of the 2023 Internal Research and Knowledge Transfer Plan of the University of Granada, and co-financed by the FEDER Andalusia Operational Programme 2021–2027.

2. “Artificial Intelligence in Europe: Rise or Decline?” (PID2023-149646NB-I00), funded by the Spanish Ministry of Science Innovation and Universities MICIU/AEI/10.13039/501100011033/ Knowledge Generation Projects 2023.

**Disclosure Statement**

The author reports there are no competing interests to declare.

**Acknowledgements**

The author reports the use of the following LLMs: ChatGPT 5.2, Gemini 3, Claude Opus 4.6, and Mistral-NeMo-12B; for the following purposes: coding assistance, chart creation, Spanish–English translation, and language improvement.

**Data Availability Statement**

The complete source code and datasets supporting the findings of this study are available at https://doi.org/10.5281/zenodo.18856716